\documentclass[prd,showpacs,preprintnumbers,twocolumn,superscriptaddress,floatfix]{revtex4-2}

\usepackage[utf8]{inputenc}
\usepackage{bm,amsmath,amssymb}
\usepackage{graphicx}
\usepackage{subfigure}
\usepackage{color}

\usepackage{scalerel}
\usepackage{tikz}
\usetikzlibrary{svg.path}
\definecolor{orcidlogocol}{HTML}{A6CE39}
\tikzset{
  orcidlogo/.pic={
    \fill[orcidlogocol] svg{M256,128c0,70.7-57.3,128-128,128C57.3,256,0,198.7,0,128C0,57.3,57.3,0,128,0C198.7,0,256,57.3,256,128z};
    \fill[white] svg{M86.3,186.2H70.9V79.1h15.4v48.4V186.2z}
                 svg{M108.9,79.1h41.6c39.6,0,57,28.3,57,53.6c0,27.5-21.5,53.6-56.8,53.6h-41.8V79.1z M124.3,172.4h24.5c34.9,0,42.9-26.5,42.9-39.7c0-21.5-13.7-39.7-43.7-39.7h-23.7V172.4z}
                 svg{M88.7,56.8c0,5.5-4.5,10.1-10.1,10.1c-5.6,0-10.1-4.6-10.1-10.1c0-5.6,4.5-10.1,10.1-10.1C84.2,46.7,88.7,51.3,88.7,56.8z};
  }
}
\newcommand\orcidicon[1]{\href{https://orcid.org/#1}{\mbox{\scalerel*{
\begin{tikzpicture}[yscale=-1,transform shape]
\pic{orcidlogo};
\end{tikzpicture}
}{X}}}}

\usepackage{hyperref}
\usepackage{multirow}
\usepackage{soul}

\hypersetup{colorlinks = true, allcolors = blue}

\newcommand{\TcXdisc}{158.7{}_{{}\mathop-2.3}^{{}\mathop+2.6}\,\mathrm{MeV}}  
\newcommand{\TpeakXus}{160\,\mathrm{MeV}}  
\newcommand{\TcXdiscHISQ}{157.4(7)\,\mathrm{MeV}}  

\begin{document}

\title{Aspects of the chiral crossover transition in (2+1)-flavor QCD with M\"obius domain-wall fermions}

\author{Rajiv V. Gavai\,\orcidicon{0000-0002-4539-2584}}
\affiliation{\href{https://ror.org/03ht1xw27}{Tata Institute of Fundamental Research}, 400005 Mumbai, India}

\author{Mischa E. Jaensch\,\orcidicon{0009-0005-3003-7263}}
\altaffiliation{\textbf{Corresponding authors:}\\\href{mailto:mjaensch@physik.uni-bielefeld.de}{mjaensch@physik.uni-bielefeld.de}, \href{mailto:rshanker@imsc.res.in}{rshanker@imsc.res.in}}

\author{Olaf Kaczmarek\,\orcidicon{0000-0002-6986-2341}}
\author{Frithjof Karsch\,\orcidicon{0000-0001-9994-5597}}
\affiliation{Fakult\"at f\"ur Physik, \href{https://ror.org/02hpadn98}{Universit\"at Bielefeld}, D-33615 Bielefeld, Germany}

\author{Mugdha Sarkar\,\orcidicon{0000-0002-3494-7249}}
\affiliation{Department of Physics, \href{https://ror.org/05bqach95}{National Taiwan University}, Taipei 10617, Taiwan}
\affiliation{Institute of Physics, \href{https://ror.org/00se2k293}{National Yang Ming Chiao Tung University}, Hsinchu 30010, Taiwan}

\author{Ravi Shanker\,\orcidicon{0009-0007-7424-3960}}
\altaffiliation{\textbf{Corresponding authors:}\\\href{mailto:mjaensch@physik.uni-bielefeld.de}{mjaensch@physik.uni-bielefeld.de}, \href{mailto:rshanker@imsc.res.in}{rshanker@imsc.res.in}}

\author{Sayantan Sharma\,\orcidicon{0000-0001-5775-4805}}
\affiliation{\href{https://ror.org/05078rg59}{The Institute of Mathematical Sciences}, a CI of Homi Bhabha National Institute, Chennai 600113, India}

\author{Sipaz Sharma\,\orcidicon{0000-0001-6916-2233}}
\affiliation{Physik Department, \href{https://ror.org/02kkvpp62}{Technische Universit\"at M\"unchen}, James-Franck-Stra{\ss}e 1, D-85748 Garching~b.~M\"unchen, Germany}

\author{Tristan Ueding\,\orcidicon{0000-0002-2440-1787}}
\affiliation{Fakult\"at f\"ur Physik, \href{https://ror.org/02hpadn98}{Universit\"at Bielefeld}, D-33615 Bielefeld, Germany}

\begin{abstract}
The nonsinglet part of the chiral symmetry in quantum chromodynamics (QCD) with two light flavors is known to be restored through a crossover transition at a 
pseudocritical temperature. However, the temperature dependence of the singlet part of the chiral symmetry and whether it is effectively 
restored at the same temperature is not well understood. Using (2+1)-flavor QCD configurations  generated using the  M\"{o}bius domain-wall 
discretization on an $N_\tau=8$ lattice, we construct suitable observables where the singlet and nonsinglet chiral symmetries are disentangled 
in order to study their temperature dependence across the crossover transition. From the peak of the disconnected part of the chiral susceptibility, 
we obtain a pseudocritical temperature $T_{pc}=\TcXdisc$ where the nonsinglet part of the chiral symmetry is effectively restored. From a calculation of 
the topological susceptibility and its temperature dependence we find that the singlet $\mathrm{U}_A(1)$ part of the chiral symmetry is not effectively 
restored at $T\lesssim 186\,\mathrm{MeV}$.
\end{abstract}


\maketitle

\section{Introduction}
Two important physical phenomena, spontaneous chiral symmetry breaking and confinement are responsible for the remarkable features in the phase diagram of strongly interacting matter described by quantum chromodynamics (QCD)~\cite{Ding:2015ona,Schmidt:2017bjt}. 
While other more exotic phases are anticipated from the study of model quantum field theories which share the same symmetries with QCD, the transition from a phase of color-singlet hadrons to a phase of quasiparticles carrying color degrees of freedom at small baryon densities~\cite{Bazavov:2017dus,HotQCD:2018pds,Burger:2018fvb,Borsanyi:2020fev} is extensively studied using first-principle lattice techniques. 
It is now established that these two phases of QCD are related by a smooth crossover~\cite{Aoki:2006we,HotQCD:2014kol,Bhattacharya:2014ara}. 
In the limiting case of massless quarks, chiral symmetry is restored through a true phase transition, and the corresponding critical 
temperature $T_c$ has been estimated~\cite{HotQCD:2019xnw,Kotov:2021rah}. However, its universality class has not been established yet. 

For two massless flavors of quarks, QCD exhibits a $\mathrm{U}_L(2) \times \mathrm{U}_R(2)$ chiral symmetry. 
Its nonsinglet subgroup $\mathrm{SU}_L(2) \times \mathrm{SU}_R(2)$ is spontaneously broken to $\mathrm{SU}_V(2)$ 
in the hadronic phase, giving rise to three pseudo-Goldstone modes: the pions, which are much lighter than the nucleons. 
The singlet $\mathrm{U}_A(1)$ subgroup of the chiral symmetry group does not correspond to an exact symmetry in QCD and 
is broken due to quantum effects arising from gauge interactions~\cite{Adler:1969gk,Bell:1969ts,Fujikawa:1979ay}. 
However, this anomalously broken symmetry is still believed to affect the nature of the chiral phase transition of QCD 
at zero baryon density with two light quark flavors~\cite{Pisarski:1983ms,Fejos:2024bgl,Giacosa:2024orp}. From renormalization group studies of model 
quantum field theories with the same symmetries as QCD, two interesting scenarios arise: 
(1)~If the $\mathrm{U}_A(1)$ symmetry is approximately restored at $T_c$, then the phase transition from the hadronic phase 
to the quark-gluon plasma phase is expected to be  of first order~\cite{Pisarski:1983ms,Butti:2003nu} or of second order in the 
$\mathrm{U}_L(2)\times \mathrm{U}_R(2) / \mathrm{U}_V(2)$ universality class~\cite{Pelissetto:2013hqa,Nakayama:2014sba}. 
(2)~If the magnitude of the $\mathrm{U}_A(1)$ symmetry breaking term is comparable to its zero temperature value at $T_c$, 
then the phase transition is of second order with $\mathrm{O}(4)$ critical 
exponents~\cite{Pisarski:1983ms,Butti:2003nu,Pelissetto:2013hqa,Nakayama:2014sba,Grahl:2013pba}. 
In such model studies the coefficient of the $\mathrm{U}_A(1)$ breaking term is a parameter; its magnitude can be estimated 
only from nonperturbative studies of QCD. 
It is an ongoing effort to determine the symmetry group near $T_c$ which defines the universality class of the chiral phase 
transition. The lattice regularization is the only practical method for this task and is studied intensively by different groups 
using different fermion discretizations and methods~
\cite{Chandrasekharan:1998yx,Aoki:2012yj,HotQCD:2012vvd,Chiu:2013wwa,Tomiya:2016jwr,Dick:2015twa,Brandt:2016daq,Petreczky:2016vrs,Aoki:2020noz, Ding:2020xlj,Aoki:2021qws, Kaczmarek:2021ser,Kaczmarek:2023bxb,Kovacs:2023vzi,Alexandru:2024tel,Giordano:2024jnc}. 

QCD thermodynamics on the lattice is typically studied using different variants of either Wilson or staggered fermion 
discretization schemes, both of which do not possess exact chiral symmetry on the lattice.
At finite lattice spacing, the Wilson Dirac operator breaks chiral symmetry explicitly, whereas the staggered fermion discretization
reduces the symmetry group to a remnant $\mathrm{U}(1)\times \mathrm{U}(1) $ part of the continuum chiral symmetry group,
which introduces additional lattice cutoff effects, the so-called taste violation effects.
In both cases, exact chiral and flavor symmetries are only recovered in the continuum limit. It is therefore important to study 
effects of chiral symmetry in QCD on the lattice using fermion discretizations where the chiral and the continuum limits are disentangled.

To address issues related to the restoration of chiral symmetry in lattice QCD calculations prior to taking a continuum limit, it 
is most suitable to use fermions with exact chiral symmetry on the lattice. 
Overlap fermions~\cite{Narayanan:1994gw,Neuberger:1997fp} are the only known lattice fermion discretization which 
has an exact chiral and flavor symmetry on the lattice. However, practical lattice QCD calculations with overlap fermions at finite 
temperature in the thermodynamic limit are numerically challenging due to the occurrence of exact zero modes~\cite{Fodor:2003bh,Fukaya:2006vs}. 
A closely related approach is the domain-wall fermion discretization~\cite{Kaplan:1992bt}, which realizes exact chiral symmetry by 
introducing an extra dimension which is infinitely large and a nontrivial defectlike potential. In lattice simulations where the extent
$L_s$ of this fifth dimension is finite, a small mixing between the left- and right-handed chiral modes is inevitable and is quantified 
by the so-called residual mass $m_\mathrm{res}$. The violation of chiral symmetry can be parametrized as $\mathrm{e}^{-m_{\mathrm{res}}L_s}/L_s$ ~\cite{RBC:2008cmd} and an exact chiral symmetry is recovered for an infinite fifth dimension, independently 
from the continuum limit of QCD. 

In this work we use a variant of the domain-wall discretization, known as  M\"{o}bius domain-wall fermions 
(MDWF)~\cite{Brower:2004xi,Brower:2012vk}, which is optimized to reduce the residual mass even on lattices 
with a moderately large extent of the fifth dimension. The chiral crossover transition in (2+1)-flavor QCD 
has been studied earlier using the MDWF discretization for physical light and strange quark masses on an $N_\tau=8$ 
lattice~\cite{Bhattacharya:2014ara}. 
Several studies have been performed on the fate of chiral symmetry near the crossover transition, using overlap fermions 
with fixed topology and MDWF in two- as well as (2+1)-flavor QCD, which report an effective restoration of the axial 
$\mathrm{U}_A(1)$ symmetry at the corresponding pseudocritical temperature~\cite{Cossu:2013uua,Tomiya:2016jwr,Aoki:2020noz,Ward:2024tdm}. 
On the other hand, studies for (2+1)-flavor QCD with domain-wall fermions in Refs.~\cite{HotQCD:2012vvd,Buchoff:2013nra} and also with 
MDWF~\cite{Bhattacharya:2014ara} have demonstrated that the $\mathrm{U}_A(1)$ symmetry remains strongly broken at this temperature, 
and more recently it has also been noted that the chiral susceptibility in the crossover region receives a sizeable contribution from 
the singlet $\mathrm{U}_A(1)$~\cite{Aoki:2021qws}. A recent study in (2+1)-flavor QCD with the HISQ action reports an effective restoration 
of the $\mathrm{U}_A(1)$ symmetry only at $T\gtrsim 180$ MeV after performing a careful continuum limit extrapolation~\cite{Kaczmarek:2023bxb}. 
It is thus imperative to revisit the problem through the use of fermions with exact chiral symmetry on the lattice, as we do in this work.

In this paper we perform a detailed study of observables sensitive to the different subgroups of chiral symmetry using (2+1)-flavor 
QCD configurations generated with dynamical MDWF at physical light and strange quark masses on an $N_\tau=8$ lattice with a large physical volume. 
By comparing our MDWF results with continuum extrapolated results obtained using the HISQ discretization scheme, we show that the MDWF discretization on $N_\tau=8$ lattices leads to results that agree well quantitatively with the continuum extrapolated HISQ results. This underlines that the physics of the chiral crossover transition is obtained in MDWF calculations even on a finite $N_\tau=8$ lattice and demonstrates the robustness of our results.

\section{Theoretical Background}
Since the $\mathrm{U}_A(1)$ subgroup of the chiral symmetry group is anomalous, there is no corresponding order 
parameter. For this reason, we focus on hadronic correlation functions integrated over spacetime, and analyze how these are sensitive to the singlet and nonsinglet part of the chiral symmetry. 
In QCD with two light quark flavors, the four possible meson correlation functions integrated over the spacetime volume are:
\begin{align}\label{eq:chi_sigma_def}
    \chi_\sigma  \,&=\, \frac{1}{2}\int \mathrm{d}^4x\, \left\langle \bar\sigma(x)\, \sigma(0) \right\rangle \\
    \chi_\delta  \,&=\, \frac{1}{2}\int \mathrm{d}^4x\, \left\langle \bar \delta^i(x)\, \delta^i(0) \right\rangle \\
    \chi_\eta    \,&=\, \frac{1}{2}\int \mathrm{d}^4x\, \left\langle \bar\eta(x)\, \eta(0) \right\rangle \\
    \chi_\pi     \,&=\, \frac{1}{2}\int \mathrm{d}^4x\, \left\langle \bar\pi^i(x)\, \pi^i(0) \right\rangle
    ~,
\end{align}
where the meson operators 
$\sigma(x) = \overline{\psi}_\ell(x)\, \psi_\ell(x)$ 
and 
$\delta^i(x) = \overline{\psi}_\ell(x)\, \tau^i\psi_\ell(x)$ 
denote the scalar singlet and isotriplet states and 
$\eta(x) = i\overline{\psi}_\ell(x)\, \gamma^5\, \psi_\ell(x)$ 
and 
$\pi^i(x) = i\overline{\psi}_\ell(x)\, \tau^i\, \gamma^5\, \psi_\ell(x)$ 
denote the pseudo-scalar singlet and isotriplet states, respectively. 
Here $\psi_\ell$ represents the field operator corresponding to the two degenerate light quark flavors with mass $m_\ell$, and $\tau$ represents the isospin operator, which for the $\mathrm{SU}(2)$ flavor group is given by the Pauli matrices.

In the chiral limit, the scalar correlators can be related to the disconnected part $\chi_\mathrm{disc}$ of the susceptibilities of the chiral condensate by means of the identity
\begin{align}
    \chi_\sigma \,=\, \chi_\delta + \frac{1}{2}\,\chi_{\text{disc}}
    ~.
\end{align}
Similarly, the pseudo-scalar susceptibilities are related to the susceptibilities of the condensate of the axial charge through
\begin{align}
    \chi_\eta \,=\, \chi_\pi - \frac{1}{2}\,\chi_{5,\text{disc}}
    ~.
\end{align}
Furthermore, the chiral Ward identities relate the single-flavor chiral condensate $\langle \bar\psi\psi \rangle_\ell$ to the isotriplet pseudo-scalar susceptibility by means of
\begin{align}\label{eqn:chiralWard1}
    \chi_\pi \,=\, \frac{\langle \bar \psi \psi\rangle_\ell}{m_\ell} 
    ~.
\end{align}
In the MDWF discretization scheme, defined on a lattice whose spatial and temporal extents are denoted by $N_\sigma$ and $N_\tau$ 
respectively, the chiral condensate and the disconnected part of the chiral susceptibility for a quark flavor $f$ are given 
by
\begin{align}
    \langle \bar \psi \psi \rangle_f 
    &=
    \frac{1}{N_\sigma^3 N_\tau}\, \big\langle \mathrm{tr}\big[ M^{-1}_f\, \partial_{m_f} M_f\big] \big\rangle
    \label{eqn:chiral-cond-def}
\\
    \chi_\mathrm{disc} 
    &=
    \frac{1}{N_\sigma^3 N_\tau}\, 
    \begin{aligned}[t]
    \Big(
        & \big\langle \mathrm{tr}\big[M^{-1}_f\, \partial_{m_f} M_f\big]^2 \big\rangle \\
        & -\big\langle \mathrm{tr}\big[M^{-1}_f\, \partial_{m_f} M_f\big] \big\rangle^2
    \Big)~,
    \end{aligned}
    \label{eqn:chiral-disc-dw-def}
\end{align}
where the trace is performed over spinor, color and the 
spacetime volume. Here, $M_f\equiv M(m_f)$ denotes the MDWF Dirac matrix with matrix elements defined as 
\begin{align}
    M_{s,s'}(m_f) &=
	(b_5\, D_W + 1)\, \delta_{s,s'}
\\\nonumber&
	+ (1-\delta_{s,1})\, (c_5\, D_W - 1)\, P_+\, \delta_{s-1,s'}
\\\nonumber&
	+ (1-\delta_{s,L_s})\, (c_5\, D_W - 1)\, P_-\, \delta_{s+1,s'}
\\\nonumber&
	- m_f\, \delta_{s,1}\, (c_5\, D_W - 1)\, P_+\, \delta_{L_s,s'}
\\\nonumber&
	- m_f\, \delta_{s,L_s}\, (c_5\, D_W - 1)\, P_-\, \delta_{1,s'}
	~,
\end{align}
for a quark mass $m_f$, with $s,s'=1,2,\cdots,L_s$, and where $P_\pm = \frac{1\pm \gamma_5}{2}$ are the left- and right-chiral projection operators, $D_W \equiv D_W(-M_\mathrm{5D})$ denotes the 4D Wilson 
Dirac operator with the domain-wall height $M_\mathrm{5D}$, and $b_5$ and $c_5$ are the M\"{o}bius parameters, which are tuned to reduce chiral symmetry 
breaking at fixed $L_s$. In the case of the HISQ discretization, the derivatives $\partial_{m_f}M_f$ in Eqs.~\eqref{eqn:chiral-cond-def} and~\eqref{eqn:chiral-disc-dw-def} are simply identity matrices with a factor of $1/4$.
The restoration of the nonsinglet part of chiral symmetry implies that~\cite{HotQCD:2012vvd,Buchoff:2013nra,Petreczky:2016vrs}
\[
    \chi_\pi = \chi_\sigma~,~\chi_\eta=\chi_\delta \quad\Rightarrow\quad \chi_{\text{disc}}=\chi_{\text{5,disc}}
    ~. 
\]  
However, $\chi_{5,\text{disc}}$ shares with the topological susceptibility $\chi_{\text{top}}$ the same quantum number 
which measures the fluctuations of the topological charge, defined in terms of the inverse of the Dirac operator $M_\ell$ for 
a single light quark flavor~\cite{Giusti:2004qd} as,
\begin{align}\label{eqn:topsusc}
\chi_{\text{top}} \,=\, \frac{T}{V}\,\langle Q^2\rangle \quad , \quad {\rm with}~~
    Q \,=\, m_\ell\, \text{tr}\big[\gamma_5\, M_\ell^{-1}\big] 
    ~.
\end{align}
Here the trace runs over spacetime, spinor and color indices. The topological tunnelings generate a finite topological 
charge on each gauge configuration and therefore lead to the violation of charge-parity symmetry. However, since QCD is a $CP$-even 
theory and physical results are obtained from averages over a large set of gauge configurations, the resulting topological charge, 
calculated after averaging over the full ensemble of configurations, vanishes and its fluctuations, given by $\chi_{\text{top}}$, 
are finite. From lattice studies in (2+1)-flavor QCD it is known that the value of the topological susceptibility is consistent 
with the prediction from chiral perturbation theory~\cite{Bonati:2015vqz} below the pseudocritical temperature and has a nontrivial 
temperature dependence up to twice its value~\cite{Petreczky:2016vrs}. At higher temperatures, {\it i.e.} beyond $2.5$ to $3$ times 
the pseudocritical temperature, its temperature dependence is well described by a dilute instanton gas approximation 
(DIGA)~\cite{Petreczky:2016vrs,Borsanyi:2016ksw,Bonati:2018blm,Burger:2018fvb}. 
Hence, when the nonsinglet part of chiral symmetry is restored, the relation
\begin{align}\label{eq:chidisc_eq_chitop}
    m_\ell^2\,\chi_{\text{disc}} = \chi_{\text{top}}
\end{align}
is exactly satisfied. This has been demonstrated within the HISQ discretization scheme in (2+1)-flavor QCD in the continuum  
limit~\cite{Petreczky:2016vrs}. Thus, the observable $m_\ell^2\,\chi_{\text{disc}}-\chi_{\text{top}}$ can be studied as a measure 
of the breaking of the nonsinglet part of chiral symmetry.

For the anomalous $\mathrm{U}_A(1)$ symmetry, an effective restoration would imply $\chi_\pi=\chi_\delta$~\cite{Shuryak:1993ee}.
Therefore, the difference between the two integrated correlation functions, $\chi_\pi-\chi_\delta$, is sensitive to the amount of 
$\mathrm{U}_A(1)$ symmetry breaking. Furthermore, when the nonsinglet part of chiral symmetry is restored, then $\chi_\pi=\chi_\sigma$, 
and as a result, the observable $\chi_\pi-\chi_\delta$ is equal to $1/2\,\chi_{\text{disc}}$. In turn, the amount of $\mathrm{U}_A(1)$ 
symmetry breaking in this phase is related to $\chi_{\text{top}}$, because $\chi_{\text{top}}=\chi_{\text{5,disc}}$. 
Since the temperature dependence of $\chi_{\text{top}}$ can be characterized using DIGA up to asymptotically high temperatures, 
the $\mathrm{U}_A(1)$ symmetry remains broken, reflecting its anomalous nature. A nontrivial temperature dependence of $\chi_{\text{top}}$ 
which cannot be explained within DIGA will thus denote a \emph{nontrivial} breaking of the singlet 
$\mathrm{U}_A(1)$ part of chiral symmetry. While the above relations are strictly valid in the continuum for two flavors 
of quarks in the massless limit, we study how well these observables, which are sensitive to the singlet and nonsinglet parts of the 
chiral symmetry, represent the physics of the chiral phase transition on a finite lattice using MDWF.

\section{Numerical Methods and Strategy}
\label{sec:numericaldetails}
In this work we have generated gauge configurations for (2+1)-flavor QCD using the MDWF discretization for fermions and the Iwasaki 
gauge action for the gauge links. To remove localized dislocations in the gauge field, which contribute to the magnitude of the residual 
mass, we weighted the fermion determinant with the dislocation suppressing determinant ratio (DSDR)~\cite{Renfrew:2008zfx}. We performed 
simulations for a fixed lattice size, where the numbers of sites along the spatial and temporal directions were set to $N_\sigma=32$ and 
$N_\tau=8$, respectively. The light ($m_\ell$) and strange ($m_s$) quark mass, the residual mass $m_\text{res}$, the extent of the fifth 
dimension $L_s$ and the M\"{o}bius parameters $b_5$ and $c_5$ were fixed following an earlier work using the same setup~\cite{Bhattacharya:2014ara}. 
The quark masses were fixed such that the pion mass has a value of $129.2(5)\,\mathrm{MeV}$ and the kaon mass is $462.5(5)\,\mathrm{MeV}$, 
with the ratio of the strange to light quark mass being $m_s/m_\ell\approx 27$ for all temperatures studied. 
The earlier study~\cite{Bhattacharya:2014ara} reported a pseudocritical temperature of $T_{pc}=155(8)(1)\,\mathrm{MeV}$;
in this work we perform a more precise determination of it.
We have also performed new optimizations within the hybrid Monte-Carlo (HMC) algorithm to achieve a speed-up in the generation of gauge 
configurations for this study. First, we have used six steps of Hasenbusch mass preconditioning, instead of five as used in the previous work, 
such that the mass ratios are spaced evenly. This accelerates the inversion of the Dirac matrix for temperatures $T\lesssim T_{pc}$, corresponding 
to the values $\beta=1.671, 1.689, 1.707$ of the strong coupling. Furthermore, for $T\gtrsim T_{pc}$, we used four integration steps per HMC trajectory 
for fermions and $32$ for the gauge links, as opposed to the previously used $48$ integration steps for the latter. With these optimizations, the time 
required to generate each gauge configuration was reduced by approximately $20\%$, with a typical acceptance rate of about $90\%$ for $T\lesssim T_{pc}$, 
and of more than $97\%$ for $T>T_{pc}$. The gauge configurations were saved after every fifth HMC step of unit length to reduce the auto-correlation between the generated 
configurations. Whenever possible, we combined in our analysis the datasets from the earlier analysis~\cite{Bhattacharya:2014ara} and this work. 
The parameters used for the generation of gauge configurations and the studied temperature range are summarized in Table~\ref{tab:gaugedetails}. 
The temperature values have been fixed using the $r_0$ scale, following the earlier work~\cite{Lin:2014tym}. The numbers of MDWF configurations generated 
at each temperature are summarized in Table~\ref{tab:mres}. We have estimated the residual mass on these configurations and then combined it with the 
earlier results~\cite{Lin:2014tym} for the same quantity, final results of which are also tabulated in Table~\ref{tab:mres}.

\begin{table}[t]
\centering
\begin{tabular}{c|cccc}
	\hline\hline
	$T$ (MeV) & $\beta$ & $m_\ell$   & $m_s$   & $b_5$ \\
	\hline
	149       & 1.671   & 0.00034 & 0.05538 & 2.5   \\
	154       & 1.689   & 0.00075 & 0.05376 & 2.5   \\
	159       & 1.707   & 0.00112 & 0.05230 & 2.5   \\
	164       & 1.725   & 0.00120 & 0.05045 & 2.5   \\
	168       & 1.740   & 0.00126 & 0.04907 & 2.2   \\
	177       & 1.771   & 0.00132 & 0.04614 & 2.0   \\
	186       & 1.801   & 0.00133 & 0.04345 & 2.0   \\
	\hline\hline
\end{tabular}
\caption{
    Input parameters used for the generation of gauge configurations on $32^3\times 8$ lattices with MDWF, in accordance with Ref.~\cite{Bhattacharya:2014ara}. 
    For all $\beta$ values, the extent of the fifth dimension was fixed to $L_s=16$, the domain-wall height to $M_\mathrm{5D}=1.8$ and the second M\"{o}bius parameter was set to $c_5 = b_5-1$.
}
\label{tab:gaugedetails}
\end{table}

We also compare results for different observables obtained by using the MDWF action to 
corresponding results obtained with the HISQ action. 
The latter have been obtained by performing calculations on (2+1)-flavor QCD configurations which were generated by the HotQCD collaboration in earlier studies~\cite{HotQCD:2018pds} using the HISQ action and the tree-level improved Symanzik gauge action.
The quark masses have been fixed for a physical strange-to-light quark mass ratio of $m_s/m_\ell=27$.
The lattice sizes used correspond to $N_\tau=8, 12, 16$ and the number of spatial sites are $N_\sigma=4\,N_\tau$. 

\break
The calculation of the chiral condensate and the disconnected part of the chiral susceptibility involves computing the 
trace of the product of the inverse of the Dirac matrix and its derivative with respect to the quark mass on each gauge 
configuration. We estimated these traces using the stochastic trace estimator method with a $\mathbb{Z}_2$ noise using 
$40$ random source vectors and a tolerance of $\epsilon = 10^{-8}$ for the associated CG inversions. The disconnected 
chiral susceptibility also contains the gauge mean of the squares of such traces, which we approximated using unbiased 
estimators.

In order to estimate the topological charge on each configuration, we have determined the number of zero modes of the 
overlap Dirac operator, which is used as a probe on the MDWF gauge ensembles.
The overlap Dirac operator has an index~\cite{Hasenfratz:1998ri} which is related to the topological features of the underlying gauge configurations if the gauge links are sufficiently smooth~\cite{Hernandez:1998et}.
The average value of the topological charge averaged over the available ensembles is consistent with zero, and we have calculated the mean fluctuation of the topological charge to estimate its susceptibility. 
The overlap Dirac operator $D_{ov}$ was realized such that on each configuration the Ginsparg-Wilson relation and the realization of the sign-function was implemented with a precision of $\lesssim 10^{-9}$. 
Furthermore, the zero modes were unambiguously identified from their chiralities: a typical zero mode always exhibits a chiral charge of $\pm 1$, with a precision of at least $10^{-4}$, whereas near-zero modes of $D^\dagger_{ov}\,D_{ov}$ come as pairs with equal and opposite chirality distinctly different from identity. 
The magnitudes of the near-zero modes, calculated within this mixed-action formalism, need to be properly renormalized and there is discussion in the literature on this topic~\cite{Tomiya:2016jwr}; however, the near-zero modes are not relevant for the study presented here and we will address this issue in a separate work.

For the estimation of the gauge average, as well as its error, we used the jackknife method, where we have determined the optimal number of jackknife bins for each observable and at each $\beta$-value according to the saturation of the jackknife error estimate.

\begin{table}[t]
\centering
\begin{tabular}{cc|c|c}
	\hline\hline
	$T$ (MeV) & $\beta$ & $N_\mathrm{traj}$ & $m_\mathrm{res}$ \\
	\hline
	149       & 1.671   & 4875  & 0.00175(1) \\
	154       & 1.689   & 3805  & 0.00120(1) \\
	159       & 1.707   & 12520 & 0.00090(1) \\
	164       & 1.725   & 6085  & 0.00068(1) \\
	168       & 1.740   & 5160  & 0.00057(1) \\
	177       & 1.771   & 6320  & 0.00043(1) \\
	186       & 1.801   & 5920  & 0.00026(1) \\
	\hline\hline
\end{tabular}
\caption{
    Number of equilibrated trajectories, generated by us using the MDWF action, after thermalization for each $\beta$ value, and the residual mass estimated on the total number of configurations.
}
\label{tab:mres}
\end{table}

\section{Results}
\subsection{Estimating the pseudocritical temperature}\label{sec:Tpc}
\begin{figure}[h]\centering
	\includegraphics[scale=0.64]{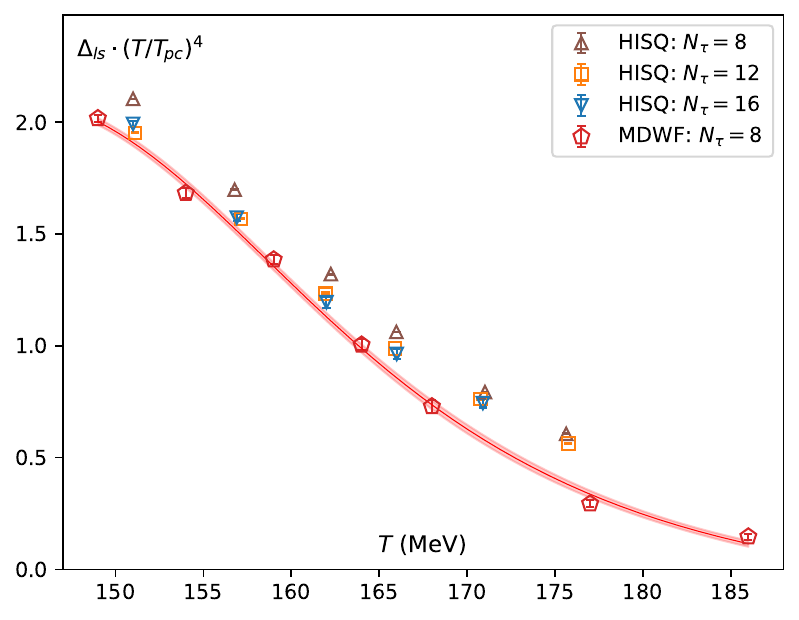}
	\caption{The subtracted chiral condensate $\Delta_{ls}$ normalized by $(T/T_{pc})^4$ is shown as a function of the temperature $T$ calculated using MDW fermions (red pentagons) on an $N_\tau=8$ lattice and compared with results from Ref.~\cite{Steinbrecher:2018jbv} obtained using the HISQ discretization on $N_\tau=8, 12, 16$ lattices. For the normalization factor $T_{pc}$ we used the pseudocritical temperature $T_{pc}=155\,\mathrm{MeV}$ reported in the earlier work~\cite{Bhattacharya:2014ara}.
	}
	\label{fig:Delta_MDWF}
\end{figure}

The chiral observables of interest are the chiral condensate and the disconnected part of the chiral susceptibility.
In order to remove part of the additive ultraviolet divergent contributions proportional to $m_\ell/a^2$ contained in the 
chiral condensate on the lattice, it is sensible to consider the weighted subtraction of the strange condensate from 
the light condensate. Further multiplying by the strange quark mass also cancels multiplicative renormalization factors. This leads to the subtracted condensate
\begin{align}
    \Delta_{ls} = \left(
        m_s\, \langle \bar\psi \psi \rangle_\ell - m_\ell\, \langle \bar\psi \psi \rangle_s 
    \right) / T^4
    ~,
\end{align}
for which we present our results in Fig.~\ref{fig:Delta_MDWF}.
Since the transition associated with the restoration of the nonsinglet $\mathrm{SU}_A(2)$ symmetry is a smooth crossover 
in QCD with physical quark masses, this observable has a rather smooth dependence on the temperature. The chiral susceptibility, 
on the other hand, is known to be more sensitive to the restoration of symmetries and, in particular, to diverge in the chiral limit. 
This also holds true for its disconnected part, whereas its connected part is expected to diverge only if the $\mathrm{U}_A(1)$ 
symmetry is restored.

\begin{figure}[h]\centering
    \includegraphics[scale=0.61]{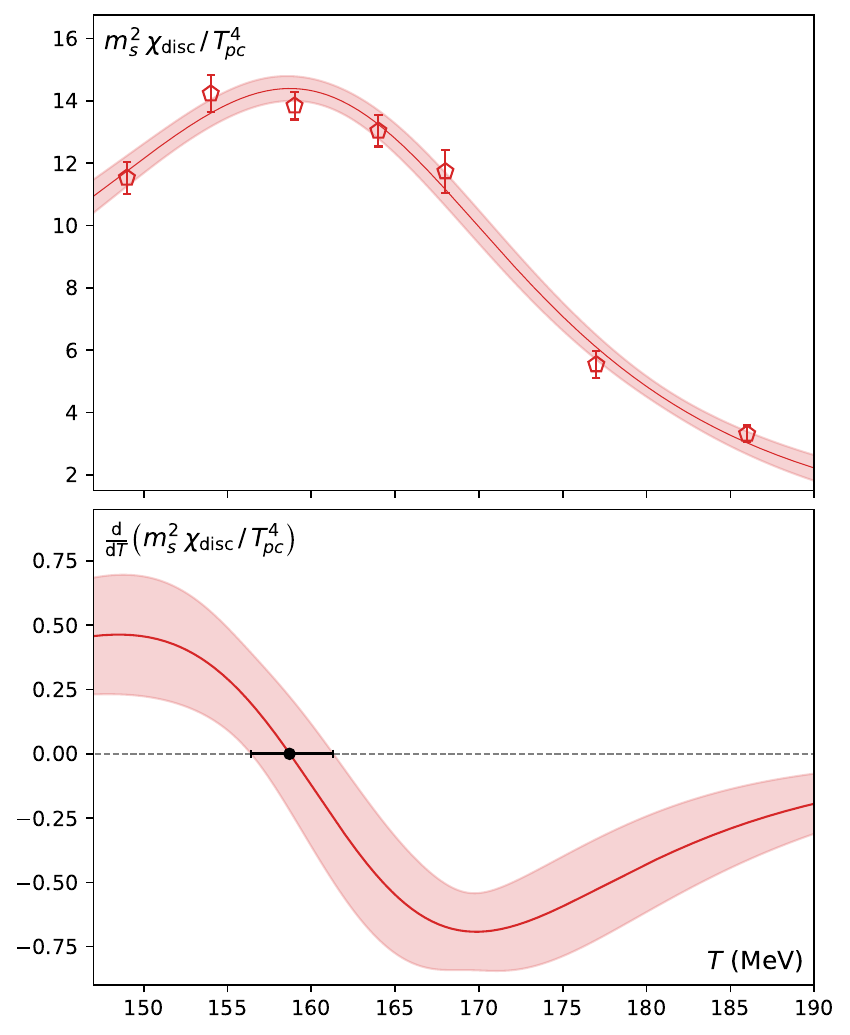}
	\caption{The disconnected part of the chiral susceptibility shown as a function of $T$ (top) and the extraction of the pseudocritical temperature $T_{pc,N_\tau=8}$ (black point) from the point where the slope of $m_s^2\chi_{\text{disc}}$ goes to zero (bottom).
	For the normalization factor $T_{pc}$ we used the pseudocritical temperature $T_{pc}=155\,\mathrm{MeV}$ reported in the earlier work~\cite{Bhattacharya:2014ara}.
}
	\label{fig:Xdisc_MDWF}
\end{figure}

The results for the disconnected part of the chiral susceptibility as a function of the temperature $T$ is shown in Fig.~\ref{fig:Xdisc_MDWF} (top). This observable shows a pronounced peak in a small range of temperatures 
and falling off both at lower and higher temperatures. The location of this peak defines a pseudocritical temperature, which we denote 
as $T_{pc,N_\tau=8}$. 

For the determination of the pseudocritical temperature, we normalized the data points of the disconnected part of the chiral susceptibility in the temperature-independent form $m_s^2\,\chi_\mathrm{disc}/T_{pc}^4$, using the temperature values determined using the $r_0$-scale and the pseudocritical temperature $T_{pc}$ from Refs.~\cite{Bhattacharya:2014ara,Lin:2014tym}, which are also given in Table~\ref{tab:gaugedetails}. The normalized data points are interpolated using a Pad\'{e} fit of order $[1/3]$. From the zero of the derivative of this rational function with respect to the temperature, shown in Fig.~\ref{fig:Xdisc_MDWF} (bottom), we have estimated the pseudocritical temperature to be 
\begin{equation}
   T_{pc,N_\tau=8} \,=\, \TcXdisc
    ~,
    \label{TcXdisc}
\end{equation}
which we henceforth refer to as $T_{pc}$.
This is consistent with the earlier estimate of the pseudocritical temperature,
$155(9)\,\mathrm{MeV}$, given in Ref.~\cite{Bhattacharya:2014ara}, 
with the improvement of reduced statistical errors.
Our estimate of $T_{pc}$ is also consistent with the corresponding ones obtained 
using the HISQ  and stout actions after performing a continuum extrapolation, which 
are $156.5(1.5)\,\mathrm{MeV}$~\cite{HotQCD:2018pds} and $158.0(6)\,\mathrm{MeV}$~\cite{Borsanyi:2020fev},
respectively. Furthermore, $\chi_\text{disc}$ becomes small only at temperatures $T> 186\,\mathrm{MeV}$, 
signalling a significant contribution from the topological fluctuations of the QCD vacuum in the vicinity 
of $T_{pc}$ which we discuss in detail in the next section. 

We have also calculated the mixed-mass susceptibility, 
\begin{align}
	\chi_{us} &=
	\begin{aligned}[t]
		& \frac{1}{N_\sigma^3 N_\tau}\,
		\Big(
		\big\langle \mathrm{tr}[M^{-1}_u\, \partial_{m_u} M_u]\, \mathrm{tr}[M^{-1}_s\, \partial_{m_s} M_s] \big\rangle \\
		& {}-{}\big\langle \mathrm{tr}[M^{-1}_u\, \partial_{m_u} M_u]\big\rangle\, \big\langle\mathrm{tr}[M^{-1}_s\, \partial_{m_s} M_s] \big\rangle
		\Big)~.
	\end{aligned}
	\label{eqn:chi-disc-mixed}
\end{align}

\begin{figure}[h]\centering
	\includegraphics[scale=0.62]{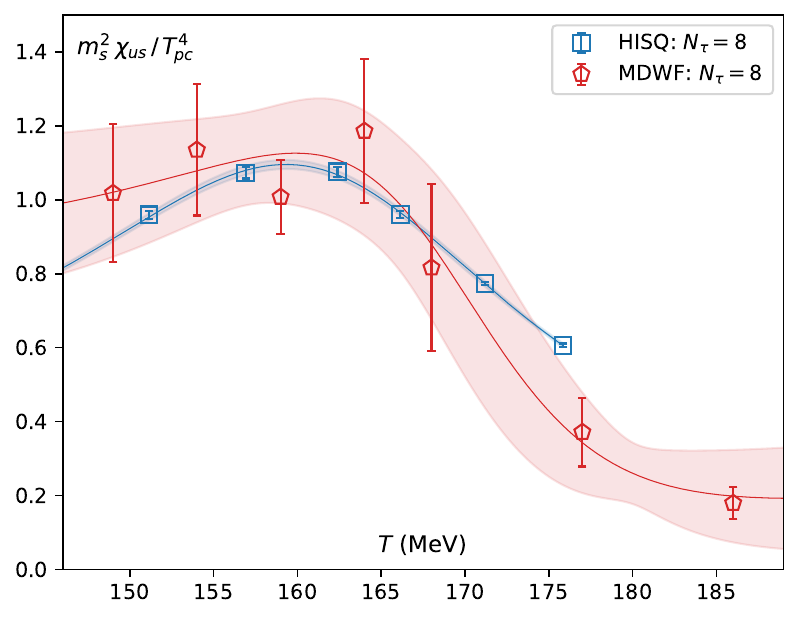}
	\caption{
	   The mixed-mass susceptibility $\chi_{us}$ as a function of the temperature $T$.
	   Shown are results from calculations with the MDWF action (red pentagons) and the HISQ action (blue squares) (taken from Ref.~\cite{Ding:2024sux}) on lattices with temporal extent $N_\tau=8$.
	   The band is obtained from a Padé fit to the MDWF data.
	   We have normalized the data with a factor of $T_{pc}^4$, using $T_{pc}=158.7\,\mathrm{MeV}$ as given in Eq.~\eqref{TcXdisc}.
	}
	\label{fig:Xus_MDWF}
\end{figure}

Results for this mixed susceptibility are shown in Fig.~\ref{fig:Xus_MDWF}. 
Our data, obtained with the MDWF action on $N_\tau=8$ lattices, are compared with corresponding results obtained in calculations with the HISQ action on lattices with
identical spatial and temporal lattice extent~\cite{Ding:2024sux}. 
We note that results obtained within the MDWF and HISQ discretization schemes are in good quantitative agreement. 
Although the statistical noise on this observable is relatively large in the MDWF calculation, it also shows a peak at $T^{\chi_{us}}_{N_\tau=8} \simeq \TpeakXus$, which is in good agreement with the quoted HISQ result $T^{\chi_{us}}_\mathrm{HISQ} = 159.39(22)\,\mathrm{MeV}$~\cite{Ding:2024sux}. 
Also the magnitudes agree well despite the different discretization errors arising in the MDWF and HISQ discretization schemes, respectively. 
As discussed in more detail in Sec.~\ref{sec:compare-light-obs}, this is quite different from the large difference observed in calculations of the disconnected chiral susceptibility on lattices with identical temporal extent, $N_\tau=8$. 
In the context of calculations with staggered fermions, this is understood to arise from taste symmetry violations which are much more important in the light quark sector than in the  strange quark sector. 
This also reaffirms the observation made in~\cite{Ding:2024sux} that the strange quark mass is relatively heavy and does not break the almost exact two-flavor chiral symmetry in the QCD action due to the light quarks explicitly. 
It therefore appears as an external parameter in the energylike scaling variable used in the scaling analysis in the vicinity of the chiral transition, just like the chemical potentials corresponding to the conserved charges.

Having discussed the restoration of the nonsinglet part of chiral symmetry, it would be also interesting to check how well the 
chiral Ward identities are followed by the MDWF action, even at finite lattice spacing.

\subsection{The fate of the singlet and nonsinglet parts of chiral symmetry}

Once the nonsinglet $\mathrm{SU}_A(2)$ part of chiral symmetry is restored at $T_{pc}$, one would then ask the question, when the singlet $\mathrm{U}_A(1)$ subgroup of the chiral symmetry is \emph{effectively} restored. We now want to estimate the magnitude of the $\mathrm{U}_A(1)$ breaking at temperatures greater than $T_{pc}$, which is done through a calculation of the observable $\chi_\pi-\chi_\delta$ as motivated earlier. According to the chiral Ward identities, the integrated pion correlator $\chi_\pi$ should be related to the chiral condensate as seen in Eq.~\eqref{eqn:chiralWard1}. On the other hand, the connected piece of the chiral susceptibility is just the scalar isotriplet correlator, $\chi_{\text{conn}} \equiv \chi_\delta$. Previous studies~\cite{HotQCD:2012vvd,Buchoff:2013nra} with domain-wall fermions have reported $\chi_\pi-\chi_\delta$ to be finite at $196\,\mathrm{MeV}$, anticipating that the $\mathrm{U}_A(1)$ symmetry will 
be effectively restored beyond this temperature~\cite{Bhattacharya:2014ara}. A different study, also with another variant of domain-wall fermions, but on relatively small lattice volumes, reports on the effective restoration of the $\mathrm{U}_A(1)$ symmetry at temperatures around $T_c$~\cite{Chiu:2023hnm}.

\begin{figure}[h]\centering
	\includegraphics[scale=0.7]{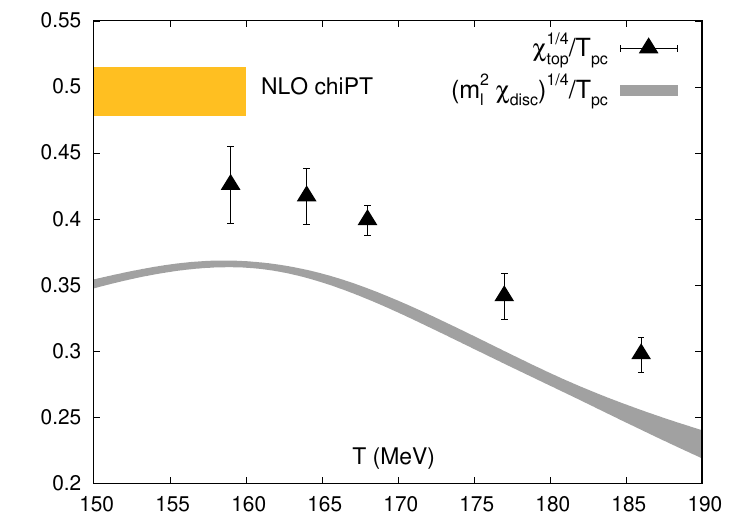}
	\caption{Comparison of our interpolated results of $(m_\ell^2\,\chi_{\text{disc}})^{1/4}/T_{pc}$ (gray band) with $\chi^{1/4}_\text{top}/T_{pc}$ (triangles) calculated in (2+1)-flavor QCD with MDWF discretization on an $N_\tau=8$ lattice
 	for temperatures above $T_{pc}$. A comparison of with the topological susceptibility calculated within 
	NLO SU(2) chiral perturbation theory (yellow band) is also shown. }
	\label{fig:chi_PT}
\end{figure}

In order to address this apparently unsettled issue, we calculate the temperature dependence of the topological susceptibility to have an independent measure of 
the $\mathrm{U}_A(1)$ breaking. We recall that the restoration of the nonsinglet part of chiral symmetry can also be inferred from Eq.~\eqref{eq:chidisc_eq_chitop}. Moreover, the finite volume correction to $\chi_\mathrm{top}$
is dominated by the $\eta'$ meson mass rather than the pion, and is expected to be mild at temperatures beyond $T_{pc}$. We obtained $\chi_{\text{top}}$ on the gauge configurations generated with MDWF, by calculating the average variance of the number of zero modes counted using as probe the overlap Dirac operator, details of which are mentioned in Sec.~\ref{sec:numericaldetails}. Our results for $\chi_{\text{top}}$ as a function of the temperature are shown in Fig.~\ref{fig:chi_PT}. From next-to-leading order (NLO) chiral perturbation theory, the topological susceptibility can be derived to be
\begin{align}\label{eqn:chitopfromchipt}
    \chi_{\text{chiPT}}^{\text{NLO}} &=
    \frac{z}{(1+z)^2}M^2_{\pi} F^2_\pi [1+\delta_1] ,\; 
\end{align}
where $\delta_1 = \frac{2M_{\pi}}{F^2_\pi}(h_1^r - h_3^r -l_4^r -l_7)$ in terms of low energy constants of SU(2) chiral perturbation 
theory~\cite{GrillidiCortona:2015jxo}. For our case, $z$, which is the ratio of mass of up and down quarks, is simply $z=1$. 
Using the latest values of $F_\pi = 94.15\,\mathrm{MeV}$, $M_{\pi} = 134.98 \,\mathrm{MeV}$ and the quantity $(h_1^r - h_3^r -l_4^r -l_7) = - 0.0114$ from the FLAG review~\cite{FlavourLatticeAveragingGroupFLAG:2021npn}, we obtain the $(\chi_{\text{chiPT}}^{\text{NLO}})^{1/4}=78.8 \pm 2.9\,\mathrm{MeV}$, where the error predominantly comes from the error in the determination of $F_\pi$. As discussed in the previous section, the contribution to the chiral observables from strange quarks is small compared to the light quarks. Hence, we do not consider three-flavor chiral perturbation theory. 
Our results for the topological susceptibility in (2+1)-flavor QCD (triangles) with MDWF start to deviate from the NLO chiral perturbation theory prediction shown as a band in Fig.~\ref{fig:chi_PT} below $T_{pc}$, alluding toward an effective restoration of the nonsinglet part of chiral symmetry.
The deviation from the chiral perturbation theory result becomes larger with increasing temperatures, showing a
$\approx 50 \%$ deviation at the highest temperature of $186\,\mathrm{MeV}$. 
However the magnitude of $\chi_{\text{top}}$ at temperatures higher than $T_{pc}$ successively gets closer to the values of $m_\ell^2\, \chi_{\text{disc}}$, besides a small deviation resulting from the finite mass of the light quarks used in this work, signalling an effective restoration of the nonsinglet part of chiral symmetry. 

Furthermore, the temperature dependence of our data for $\chi_{\text{top}}^{1/4}/T_{pc}$ at $T>160\,\mathrm{MeV}$ can be characterized by $(T/T_{pc})^{-\alpha}$ with $\alpha=2.8(1)$. 
The exponent $\alpha \approx 2$ describes the temperature dependence of the topological susceptibility in QCD with two light quark flavors within a dilute instanton gas model, which assumes fluctuations of isolated uncorrelated instantons, and should persist at asymptotically high temperatures. The fact that the exponent $\alpha > 2$, reflects a nontrivial breaking of the $\mathrm{U}_A(1)$ symmetry up to $T\approx 186\,\mathrm{MeV}$, which is unlikely due to a dilute instanton gaslike scenario. It is also known from a recent study with the HISQ discretization that the $\mathrm{U}_A(1)$ symmetry is effectively restored at $T \gtrsim 180 \,\mathrm{MeV}$,  after performing a continuum extrapolation~\cite{Kaczmarek:2023bxb}. 
This is consistent with the fact that we do not observe an effective restoration of the $\mathrm{U}_A(1)$ part of chiral symmetry up to the highest temperature we have studied using MDWF, which is $T=186\,\mathrm{MeV}$.

\subsection{Comparing chiral observables in the light quark sector calculated with MDWF and HISQ}\label{sec:compare-light-obs}
Having obtained precise results on the chiral observables using MDWF, even on a finite $N_\tau=8$ lattice, it is imperative to determine how well these results compare with results obtained using different fermion discretizations.
The staggered fermion discretization scheme explicitly breaks the nonsinglet flavor symmetries on the lattice and only respects a $\mathrm{U}(1)$ subgroup of the whole chiral symmetry group of QCD with two light quark flavors, which is an admixture of the physical flavor and the taste symmetries. 
It is known that the chiral symmetries in terms of the chiral Ward identities are recovered in the continuum limit for observables calculated using the HISQ discretization. 
This has been shown, for example, through the exact matching of the disconnected part of the chiral susceptibility and the topological susceptibility in the continuum limit for physical gauge ensembles in (2+1)-flavor QCD generated with the HISQ action~\cite{Petreczky:2016vrs,Kaczmarek:2023bxb}. 
On the other hand, in earlier work with MDWF in Ref.~\cite{Bhattacharya:2014ara}, it was observed that $\chi_\text{disc}$ calculated using domain-wall fermions was $50\%$ larger than the result computed using the staggered discretization in the crossover region. 
It would thus be important to now address this puzzle, especially since we know that for a meaningful comparison one has to perform a continuum extrapolation of the staggered results. 

\begin{figure}[th]\centering
    \includegraphics[scale=0.65]{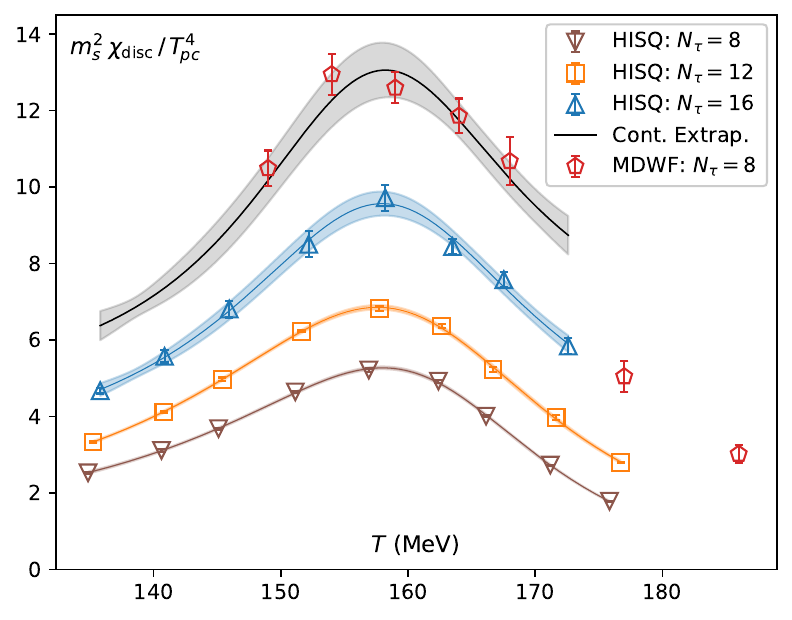}
	\caption{$m_s^2\, \chi_\mathrm{disc}/T_{pc}^4$ computed with the MDWF action for $N_\tau=8$, in comparison to results obtained using the HISQ action. The continuum estimate was computed from the HISQ $N_\tau=12,16$ results only.
    For the normalization factor $T_{pc}$ we used the pseudocritical temperature $T_{pc}\equiv  T_{pc,N_\tau=8}$ given in Eq.~\eqref{TcXdisc}.
	}
	\label{fig:chidisccomp}
\end{figure}

Results for the observable $\chi_\text{disc}$ as a function of the temperature are shown in Fig.~\ref{fig:chidisccomp}, comparing the results obtained using the HISQ discretization for lattice spacings corresponding to $N_\tau=8,12, 16$ with our results calculated using MDWF on an $N_\tau=8$ lattice.
From a continuum extrapolation of the peak positions associated with the HISQ $N_\tau=8,12, 16$ data for $\chi_\mathrm{disc}$, we have estimated the pseudocritical temperature with an $\mathcal{O}(1/N_\tau^2)$ fit resulting in $T_{pc}^\mathrm{HISQ}=\TcXdiscHISQ$, which agrees well within errors with our corresponding MDWF $N_\tau=8$ 
estimate of $T_{pc,N_\tau=8}=\TcXdisc$, as given in Eq.~\eqref{TcXdisc}.
Unlike the location of its peaks, the numerical values of $\chi_\mathrm{disc}$  obtained using HISQ action on an $N_\tau=8$ lattice are clearly not within the $\mathcal{O}(a^2)$ scaling regime.
We thus excluded these results in determining the continuum estimate of  $\chi_\mathrm{disc}$, thereby using only the $N_\tau=12, 16$ results. The fact that this difference in the scaling behavior shows up in the magnitude 
of the disconnected chiral susceptibility and is not reflected in the location of its peak is
not unfamiliar and shows up also in other fluctuations observables, e.g. cumulants of 
conserved charge fluctuations. Here it can be
addressed to large cutoff effects arising
from 
taste violations inherent to staggered fermion 
actions. This may also be the case for the
disconnected chiral susceptibility as it is known
that in particular at low temperatures the 
chiral condensate is dominated by contributions from the pseudo-scalar meson sector \cite{Biswas:2022vat}, where taste violation effects are large.
Furthermore, the data for $m_s^2\, \chi_\mathrm{disc}/T_c^4$ for different lattice cutoffs obtained using HISQ 
discretization thus shows that taste violation effects strongly influence chiral symmetry breaking at finite lattice spacing,
resulting in the presence of large cutoff effects in observables sensitive to the pseudo-scalar meson sector of QCD.
The comparison between different fermion discretizations also highlights the importance of calculating chiral observables using lattice fermion discretizations, 
e.g. MDWF, where the chiral symmetry breaking and finite cutoff effects are largely disentangled.

Furthermore, our MDWF results for $\chi_\mathrm{disc}$ lie within the continuum estimate band obtained 
using the HISQ data, even though our calculations with MDWF were performed on a finite cutoff i.e., an 
$N_\tau=8$ lattice. Of course, results for physical observables obtained by using different fermion 
discretizations should agree in the continuum limit. The fact that our results obtained using MDWF on 
a finite lattice spacing already agree with the continuum estimate of the HISQ results suggests that finite 
cutoff effects are subdominant for finer lattices, i.e., for $N_\tau >8$. 
In fact, from Eq.~\eqref{eqn:chiral-disc-dw-def} it is evident that the additional term contained in the mass derivative of the MDWF operator becomes irrelevant on finer lattices, as its magnitude decreases with decreasing the lattice spacings. Nonetheless, in order to finally arrive at a well controlled comparison between different discretization schemes this should eventually be verified explicitly in a direct lattice QCD calculation within the MDWF discretization scheme using larger temporal lattice extents, $N_\tau$.

\begin{figure}[h]\centering
	\includegraphics[scale=0.67]{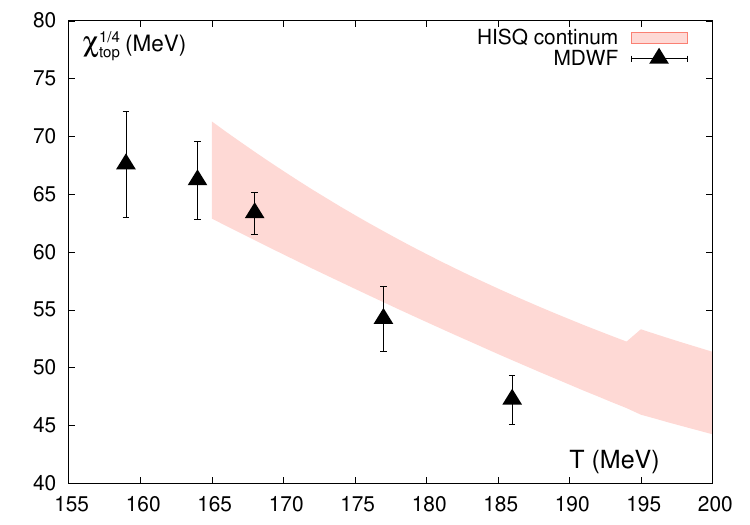}
	\caption{Comparison of $\chi_\mathrm{top}^{1/4}$ (in MeV) between continuum extrapolated results using the HISQ discretization taken 
	from Ref.~\cite{Petreczky:2016vrs} versus the values we have obtained in (2+1)-flavor QCD using the MDWF discretization.}
	\label{fig:chitdw-hisq}
\end{figure}

Next, we compare how well the topological tunnelings that occur in QCD are represented in the gauge configurations generated using the 
HISQ and MDWF discretizations for fermions. This comparison is important for two reasons. First, for the domain-wall fermions, one can 
be assured that the topological tunnelings are adequately represented even at non-vanishing lattice spacing without any significant 
contributions arising from localized dislocations whose size is of the order of a single lattice spacing, quite unlike in the case for 
staggered fermions at a similar value of the lattice cutoff. 
Second, the domain-wall Dirac matrix violates continuum chiral symmetries through terms which are irrelevant 
in the continuum limit, unlike the HISQ discretization where the chiral symmetry group and hence the notion of an index is altered 
on a finite lattice due to rooting. It is not apriori clear what are the consequences of this fact for the topological 
susceptibility obtained using the HISQ discretization, i.e., how fine lattices would be required to perform a correct continuum 
extrapolation of this observable. In the literature there have been extensive discussions on whether the topological fluctuations 
are correctly accounted for in rooted staggered fermions~\cite{Creutz:2007yg}. In the weak-coupling lattice perturbation theory it 
has been shown that a four flavor-taste staggered determinant factors into four equivalent determinants and that the correct flavor 
symmetries emerge in the continuum limit~\cite{Durr:2005ax,Sharpe:2006re,Kronfeld:2007ek,Golterman:2008gt}. 
For staggered fermions, however, the flavor singlet part of chiral symmetry which is sensitive to topology, has to be defined 
carefully on a finite lattice. Previous studies reconstructing the t'Hooft vertex and the flavor singlet meson correlator from 
the near-zero-mode quartets, showed the emergence of the expected behavior of these observables in the continuum~\cite{Donald:2011if} 
and more recently reproduced the $\eta'$ meson mass using the staggered discretization of QCD in the continuum 
limit~\cite{Verplanke:2024msi}.

The $\chi_{\text{top}}^{1/4}$ calculated on the gauge configurations generated with MDWF discretization on an $N_\tau=8$ lattice, 
is compared with corresponding continuum extrapolated values obtained by using the HISQ discretization from Ref.~\cite{Petreczky:2016vrs} 
in Fig.~\ref{fig:chitdw-hisq}. The topological charge on the HISQ configurations was obtained using the improved clover definition, 
after performing Symanzik flow to reduce the ultra-violet noise resulting from gauge fluctuations~\cite{Mazur:2021zgi}. We observe 
that, even though the results obtained from the MDWF ensembles are for a fixed lattice spacing, these are consistent with the continuum 
extrapolated results obtained using HISQ fermions. It is to be noted that the calculations involving MDWF consisted of $\mathcal{O}(10^2)$ 
i.e., only about a tenth of the configurations we have generated, whereas the HISQ results were obtained using $\mathcal{O}(10^3)$ 
configurations. Our results demonstrate a first comparison of the topological susceptibility, at finite temperature, of continuum 
extrapolated HISQ results with MDWF, highlighting the importance of improved realization of chiral symmetry on the lattice also 
for a proper topological sampling among the gauge configurations.

\section{Summary and Outlook}

In this paper we have shown the importance of the MDWF discretization, which respects the chiral symmetry to a good extent on a finite lattice, for understanding the thermodynamics of the chiral crossover transition in QCD. Since the light quark mass is significantly smaller than any other scale in finite temperature QCD describing physical quarks, it is sensible to revisit the question whether the singlet and nonsinglet subgroups of chiral symmetry are effectively restored simultaneously.  Addressing this question on the lattice using the HISQ discretization is more subtle, since it breaks the nonsinglet chiral symmetry group to a U(1) subgroup, as a result of a mixing between the spin-flavor degrees of freedom. Further, due to lack of an exact index of the HISQ Dirac matrix, the topological fluctuations, which are responsible for the \emph{effective} restoration of the singlet part of the chiral symmetry in the continuum, cannot be explicitly related to each other on the lattice. While taking the continuum limit of chiral observables such as  $\chi_\text{disc}$ calculated using the HISQ discretization is strictly necessary to extract the corresponding pseudocritical temperature, we have shown that the MDWF discretization reproduces the continuum QCD results well already on a finite $N_\tau=8$ lattice. It is therefore expected that calculating $\chi_\text{disc}$ using MDWF on a finer lattice, would show only a weak cutoff dependence, since the chiral and continuum limits are very well disentangled, in contrast to staggered fermions. Furthermore, we find that the HISQ results for $\chi_\text{disc}$ on an $N_\tau=8$ lattice have large cutoff effects and cannot be taken into consideration in the process of estimating its continuum values using an $\mathcal{O}(a^2)$ extrapolation. We have also shown that the topological fluctuations quantified by the observable $\chi_\text{top}$, obtained from calculations on the $N_\tau=8$ MDWF gauge ensembles, are similar in magnitude to the continuum extrapolated values calculated on the HISQ gauge ensembles. This highlights once more the role of chiral symmetry in also realizing the topological fluctuations more accurately.

By determining the exponent $\alpha$ controlling the temperature dependence of $\chi_\text{top}$ using MDWF calculations, we have shown that at $T \approx 185\,\mathrm{MeV}$ the singlet part of the chiral symmetry is not yet \emph{effectively} restored, i.e., to its asymptotically high temperature value given by a dilute instanton gas approximation.
This finding suggests that the chiral phase transition in the massless limit of (2+1)-flavor QCD is of second order with critical exponents belonging to the $\mathrm{O}(4)$ universality class.

Apart from extracting the pseudocritical temperature $T_{pc} = \TcXdisc$ from the peak of 
the disconnected chiral susceptibility $\chi_\text{disc}$, we independently show that the nonsinglet part of chiral symmetry is 
also \emph{effectively} restored in the same temperature range. This is evident from the deviation of $\chi_\text{top}^{1/4}/T_{pc}$ from the chiral perturbation theory 
prediction and from the realization of the relation $\chi_\text{top}=m_\ell^2 \chi_\text{disc}$ up to $\mathcal{O}(m_\ell^4)$ corrections due to the
finite mass of the light quarks.

While it is anticipated that the cutoff dependence of chiral observables, such as $\chi_\mathrm{disc}$, computed using the MDWF 
discretization is much weaker compared to calculations using the HISQ action, which suffer from taste violation effects, it would 
nonetheless be instructive to explicitly verify this hypothesis by performing our computations and analysis on a finer  
lattice. It would also be interesting to directly calculate the scalar isotriplet susceptibility through precise computation of the 
connected chiral susceptibility. This would require additional inversions of the Dirac matrix. Furthermore, it would be important to 
study in detail the properties of the eigenvalues and eigenvectors~\cite{Lin:2014tym,Sharma:2016cmz,Aoki:2020noz,Pandey:2024goi} 
of the MDWF operator in order to better understand the mechanism underlying the chiral symmetry restoration from the temperature 
dependence of the eigenvalue density.

\section*{Acknowledgments}
This work was supported by the Deutsche Forschungsgemeinschaft
(DFG, German Research Foundation) Proj. No. 315477589-TRR 211; 
the PUNCH4NFDI consortium
supported by the Deutsche Forschungsgemeinschaft (DFG, German Research Foundation) with project number 460248186 (PUNCH4NFDI). M..S acknowledges support by the Taiwanese NSTC Grant No. 113-2639-M-002-006-ASP.
The authors gratefully acknowledge the Gauss Centre for Supercomputing e.V.~\cite{GaussCentre} for funding this project by providing computing time through the John von Neumann Institute for Computing (NIC) on the GCS Supercomputer JUWELS at J\"{u}lich Supercomputing Centre (JSC).
Additional computations have been performed on the GPU Cluster at Bielefeld University and at the Institute of Mathematical Sciences, Chennai. Our codes were developed in part using the \texttt{Grid Python Toolkit (GPT)}~\cite{Lehner:GPT} which uses the \texttt{Grid} library~\cite{Boyle:2016lbp}. Part of the data analysis has been performed using routines from the \texttt{AnalysisToolbox}~\cite{Altenkort:2023xxi} code developed by the HotQCD Collaboration.
Part of the HISQ calculations were performed using the \texttt{SIMULATeQCD}~\cite{Mazur:2021zgi,HotQCD:2023ghu} library. We also thank Christoph Lehner for his advice and help in using \texttt{GPT}, and Yu Zhang for helpful
discussions on DWF calculations.

\section*{Data Availability}
The data that support the findings of this work are openly available~\cite{DataPub}.

\bibliography{dwpaper.bib}

\end{document}